\documentclass[5p]{elsarticle}

\usepackage{lineno,hyperref,color}
\modulolinenumbers[5]

\journal{Astroparticle Physics}
\usepackage{amsmath}
\usepackage{multirow} 
\usepackage{morefloats}  
\usepackage{textcomp} 
\biboptions{sort&compress}
\usepackage[utf8]{inputenc}
\usepackage{dcolumn}
\newcolumntype{d}[1]{D{.}{.}{#1}}

\graphicspath{{../Figures/}}

\begin{document}
	\begin{frontmatter}	
		\title{The muon intensity in the Felsenkeller shallow underground laboratory}	
		
		\author[a,b]{F. Ludwig}
		\author[a,b]{L. Wagner}
		\author[a,c]{T. Al-Abdullah}
		\author[d]{G. G. Barnaföldi}
		\author[a]{D. Bemmerer}
		\ead{d.bemmerer@hzdr.de}
		\author[f]{D. Degering}
		\author[a,b]{K. Schmidt}
		\author[e]{G. Surányi}
		\author[a]{T. Szücs}
		\author[b]{K. Zuber}
		
		\address[a]{Helmholtz-Zentrum Dresden-Rossendorf (HZDR), Bautzner Landstr. 400, 01328 Dresden, Germany}
		\address[b]{Technische Universität Dresden (TU Dresden), 01069 Dresden, Germany}
		\address[c]{Hashemite University, Physics Department, 13133 Zarqa, Jordan}
		\address[d]{Wigner Research Centre for Physics of the Hungarian Academy of Sciences (MTA Wigner),  H-1525 Budapest, Hungary}
		\address[e]{MTA-ELTE Geological, Geophysical and Space Science Research Group, H-1117 Budapest, Hungary}
		\address[f]{VKTA -- Strahlenschutz, Analytik \& Entsorgung Rossendorf e.V., 01328 Dresden, Germany}
		
		\begin{abstract}
			The muon intensity and angular distribution in the shallow-underground laboratory Felsenkeller in Dresden, Germany have been studied using a portable muon detector based on the close cathode chamber design. Data has been taken at four positions in Felsenkeller tunnels~VIII and IX, where a new 5\,MV underground ion accelerator is being installed, and in addition at four positions in Felsenkeller tunnel~IV, which hosts a low-radioactivity counting facility. At each of the eight positions studied, seven different orientations of the detector were used to compile a map of the upper hemisphere with 0.85$^\circ$ angular resolution. The muon intensity is found to be suppressed by a factor of 40 due to the 45\,m thick rock overburden, corresponding to 140 meters water equivalent. 
			The angular data are matched by two different simulations taking into account the known geodetic features of the terrain: First, simply by determining the cutoff energy using the projected slant depth in rock and the known muon energy spectrum, and second, in a Geant4 simulation propagating the muons through a column of rock equal to the known slant depth. The present data are instrumental for studying muon-induced effects at these depths and also in the planning of an active veto for accelerator-based underground nuclear astrophysics experiments.
		\end{abstract}
		
		\begin{keyword}
			Muon intensity \sep Underground laboratories \sep Geant4 \sep nuclear astrophysics \sep wire chambers \sep muon tomography \sep muon radiography
		\end{keyword}
	\end{frontmatter}
	
	
	\section{Introduction}
	\label{sec:Introduction}
	
	The ambient muon intensity at the Earth's surface is caused by primary cosmic ray interactions in the upper atmosphere \cite{Grieder01}. Since cosmic-ray induced muons are directly ionizing particles and, due to their high energy, very penetrative, they may be used in imaging and tomography applications, such as looking for hidden chambers in pyramids \cite{Alvarez70,Morishima17} or hollow spaces inside volcanos \cite{Nagamine95,Tanaka14,Olah18}, scanning for concealed nuclear material \cite{Borozdin03}, imaging the damaged reactor cores of Fukushima \cite{Kume16,Morris14,Borozdin12}, or monitoring stored carbon dioxide in geological reservoirs \cite{Klinger15}.
	
	However, because of their ubiquity and high energy, in low-background radiation measurements muons may induce backgrounds, either by direct ionization or by the production of secondary particles, that are easy to attenuate but difficult to completely suppress. Therefore, ultra-low background laboratories are frequently placed in underground settings, where due to the energy loss in many meters of rock overburden the muon intensity is strongly suppressed \cite{Heusser95,Formaggio04}.
		
	Possible applications for such underground laboratories are ultra-low level $\gamma$-ray spectroscopy \cite{Laubenstein04,Niese98,Hult00,Theodorsson03,Povinec04,Pellicciari05,VanBeek12,Sivers14,Heusser15}, ultra-pure material development \cite{Aalseth12}, or nuclear astrophysics \cite{Broggini18-PPNP,Bemmerer18-SNC}.
	
	When the rock overburden exceeds 1000\,m thickness, the muon intensity is suppressed by six orders of magnitude or more, so that it usually does not limit experiments any more. In these deep-underground settings, solar neutrino flux measurements \cite{SNO10-PRC,SuperK11-PRD,Borexino14-Nature}, dark matter searches \cite{Xenon18-PRL}, and rare event studies \cite{Legend17-AIPCP} are possible, and also nuclear astrophysics greatly benefits \cite{Broggini18-PPNP,Trezzi17-APP}. 
	
	The present work reports on a study of the muon intensity in the Felsenkeller shallow-underground laboratory in Dresden, Germany, using experiments, calculations, and Monte Carlo simulations. This site is sufficiently deep underground that other cosmic-ray induced effects besides muons can be neglected, and sufficiently shallow that the muon intensity is high enough for a detailed study in a reasonable time frame. 
	
	The muon data developed here will be used in forthcoming work \cite{Grieger19-inprep} addressing the recent debate on muon-induced neutrons underground \cite{Du18-APP}. 
In addition, it will be instrumental for designing an active veto for underground nuclear astrophysics experiments planned in Felsenkeller tunnels VIII and IX \cite{Bemmerer18-SNC}. It has been shown previously  \cite{Szucs12-EPJA,Szucs15-EPJA} that such a veto may reduce the observed background in $\gamma$-detectors typical for in-beam nuclear astrophysics experiments to a level that is close to the background in the same detectors deep underground, underlining the importance of a proper muon veto.
		
	\begin{figure*}[t!!]
		\centering
		\includegraphics[width=\textwidth,trim=7.7mm 8.5mm 20mm 9.1mm,clip]{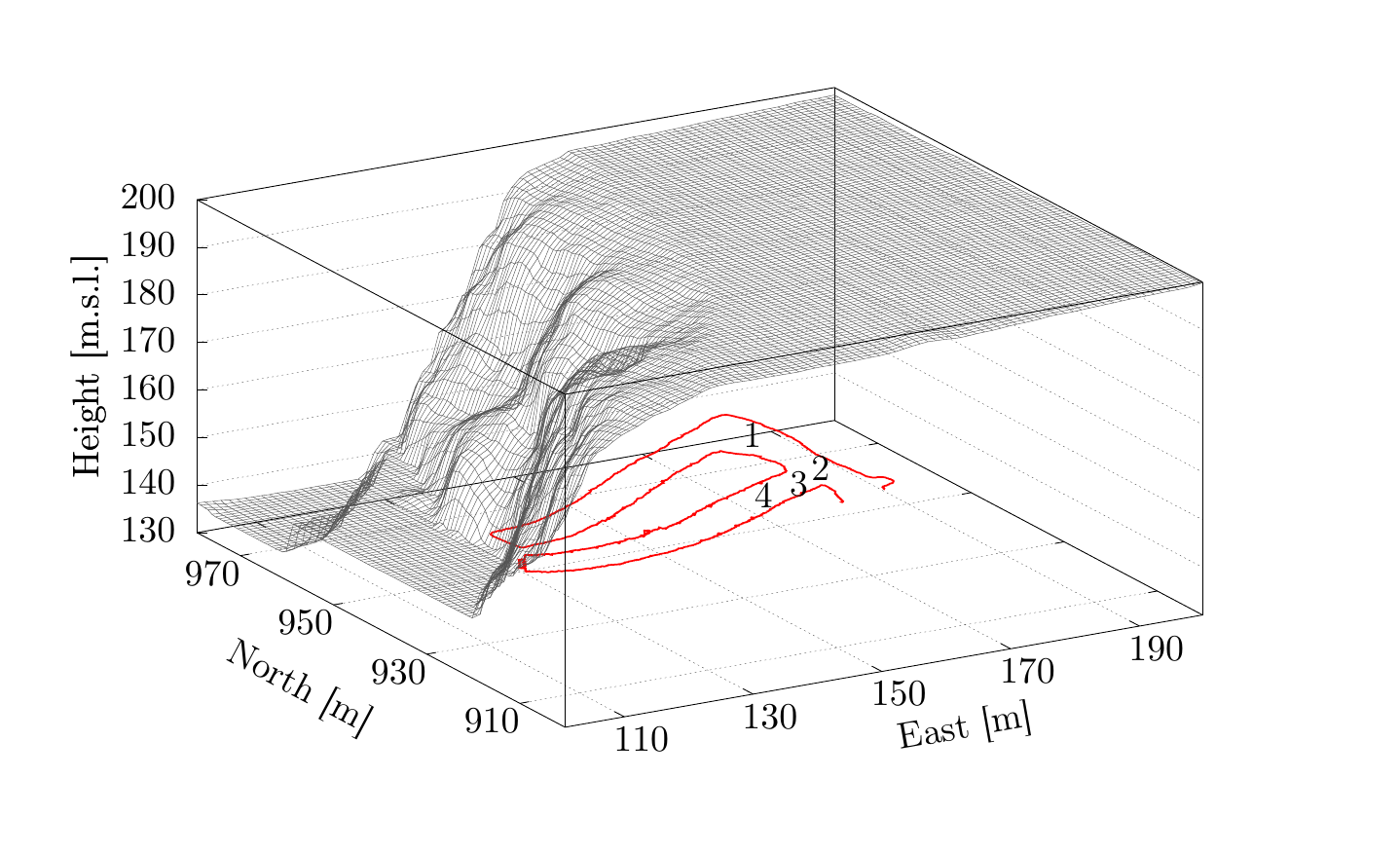}
		\caption{Terrain map above Felsenkeller tunnels~VIII and IX, as given by the DGM1 geodata. The distances given on the axes are relative to the reference point (5409000,5654000) in the RD/83 system. The flat plateau above the tunnels, the sharp cliff edge, and the flat valley of the river Weißeritz are all well visible. Measurement points 1-4 are indicated by arabic numbers. Tunnel IV with measurement points 5-8 is slightly further south and not shown on the map. See text for details.}
		\label{fig:Terrain}
	\end{figure*}
		
	This work is organized as follows. The underground site is described in sec.~\ref{sec:Felsenkeller}. Section~\ref{sec:Experiment} introduces the experimental setup, including the REGARD muon telescope used here, and experimental procedures. The data analysis and results are presented in sec.~\ref{sec:Analysis}. The data are then matched, first, by a calculation based on the known range-energy relation, and second, by a Monte Carlo simulation using the Geant4 framework (sec.~\ref{sec:Calculation}). A discussion is offered in sec.~\ref{sec:Discussion}. The conclusions, a summary and an outlook are given in sec.~\ref{sec:Summary}. 
	
	\section{Description of the underground site studied}
	\label{sec:Felsenkeller}
	
	\begin{figure}[tb]
		\centering
		\includegraphics[width=\columnwidth,trim=20mm 8mm 8mm 10mm,clip]{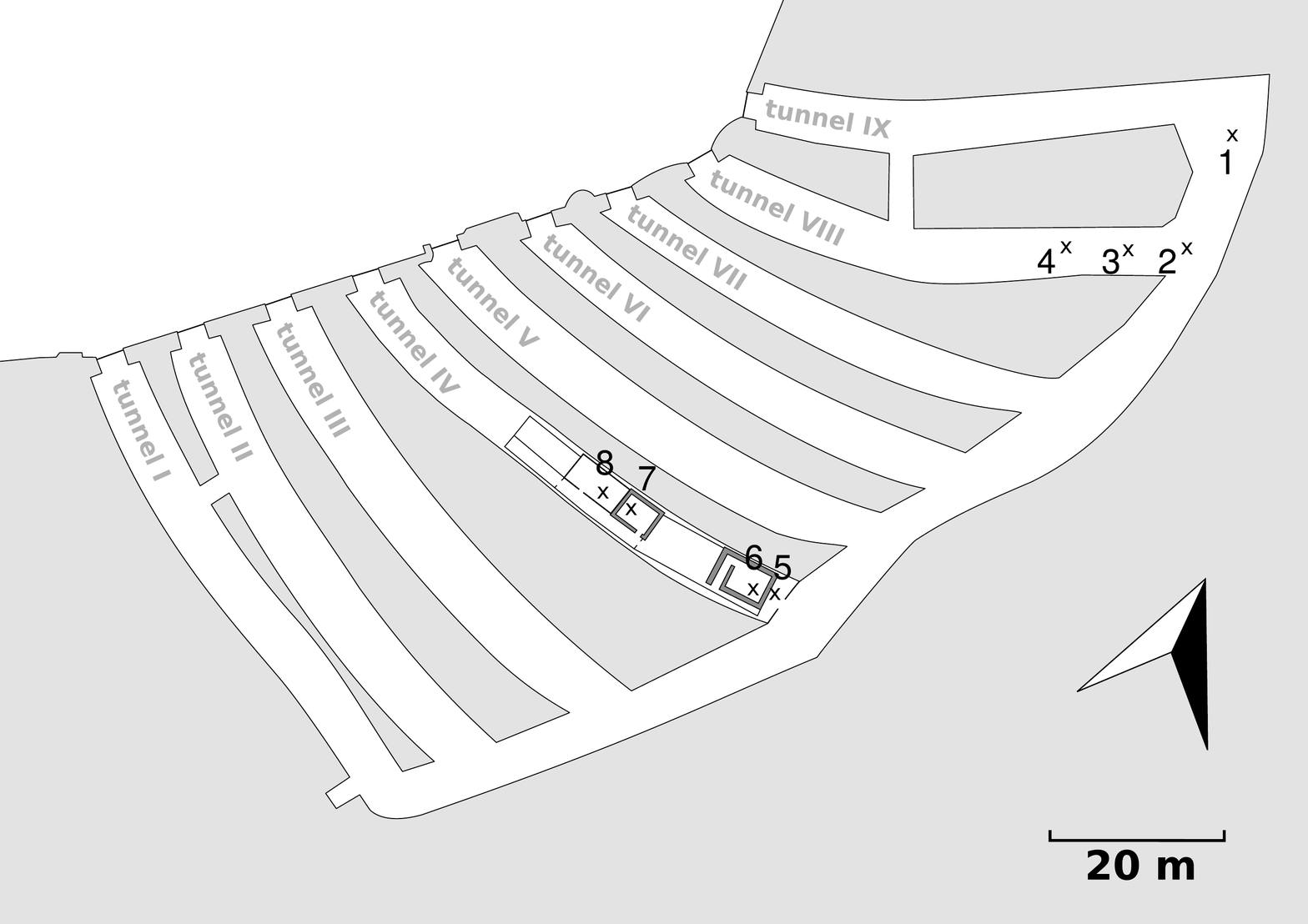}
		\caption{Map of the Felsenkeller underground site. Measurement positions~1-4 are located in tunnels~VIII and IX, and measurement positions 5-8 in tunnel~IV.}
		\label{fig:Map}
	\end{figure}
	
	The shallow-underground site Felsenkeller is located in the Plauenscher Grund district, inside the city of Dresden, Germany. The site extends along the Weißeritz river, a tributary of the Elbe, and was used as a quarry until the 18th century, then converted to a brewery, which in turn closed in 1991. The terrain is characterized by a steep cliff that runs from Northeast to Southwest, an approximately flat high plain at 200\,m a.s.l. and a river floodplain at 140\,m a.s.l. (Fig.~\ref{fig:Terrain}).
	
	Nine horizontal storage tunnels were dug into the rock from 1856 to 1859. All nine tunnels have horizontal access and are interconnected in a comb-like structure (Fig.~\ref{fig:Map}). The site is protected from cosmic rays by an overburden of 45\,m of hornblende monzonite rock, part of the "Meißner Massiv" formation. The density of rock samples taken from the tunnels~VIII and IX was here found to be (2.69$\pm$0.06) g/cm$^3$. The hornblende monzonite contains a thin sandstone layer, here neglected for simplicity.

The tunnels are on the level of the river valley, 140\,m above sea level.
	
	The present study was motivated by the project to install an ion accelerator and a low background activity-counting facility in Felsenkeller tunnels~VIII and IX \cite{Bemmerer18-SNC}. Consequently, the muon intensity was studied in four places that are located in these two tunnels (Fig.~\ref{fig:Map}):
	\begin{enumerate}
		\item Position 1 is located deep in the connection tunnel behind tunnels~VIII and IX.
		\item Position 2 lies at the very end of tunnel~VIII.
		\item At position 3, after the present measurements were concluded, the in-beam measurement bunker was installed  \cite{Bemmerer18-SNC}.
		\item Position 4 marks the activation measurement bunker, installed after the present measurements were concluded \cite{Bemmerer18-SNC}.
	\end{enumerate}
	
	In addition, four more places were studied in tunnel~IV, which hosts a $\gamma$-counting facility established in 1982 \cite{Helbig84-Isotopenpraxis} and enlarged in 1995 \cite{Niese96-Apradiso}:
	\begin{enumerate}
		\setcounter{enumi}{4}
		\item Position 5 at the very end of tunnel~IV. 
		\item Position 6 is located in Messkammer 1 (hereafter called MK1). This bunker is shielded by 68~cm serpentinite rock and 2~cm old (i.e. pre-1945) steel for a total areal density of 200~g/cm$^2$ \cite{Helbig84-Isotopenpraxis}. 
		\item Position 7 is located in Messkammer 2 (hereafter called MK2). MK2 has 210~g/cm$^2$  areal density shielding, made up of 6~cm old steel, 3~cm lead and 27~cm iron pellets \cite{Niese96-Apradiso}. 
		\item Finally, position 8 lies in a workshop area (hereafter called WS), shielded from the surrounding hornblende monzonite rock just by a thin composite wall to allow climatization of the inner area.
	\end{enumerate}
	
	For reference purposes, a ninth position was studied atop the rock burden, approximately vertically above positions 1-4, inside the "Hoher Stein" observation tower.
	
	\section{Experiment}
	\label{sec:Experiment}
	
	
	\subsection{Description of the muon detector used}
	
	For the measurements, a muon detector based on the close cathode chamber (CCC) design by the Hungarian REGARD group was used. This device was constructed to combine a large effective detector area, high robustness and satisfactory angular resolution with transportability, low power consumption and cost efficiency \cite{Varga11,Barnafoeldi12,Varga13}. It is henceforth called the REGARD muon telescope and was used previously in some shallow underground locations in Hungary \cite{Barnafoeldi12}.
	
	The muon telescope hosts six chambers of 25.6$\times$25.6 cm$^2$ active area that are spaced vertically by 3.5\,cm in a common plexiglass box. The working gas is ATAL (82\% argon and 18\% carbon dioxide) at atmospheric pressure, continuously flushing with typically 0.5\,l/h. Each of the chambers is mounted on a printed circuit board that is divided into 64 pads, each 4\,mm wide. 1.5\,mm above the ground plate, each chamber contains 64 field wires, that are spaced by 4\,mm at a working voltage of -600\,V. In the center of two field wires are sense wires on a potential of 1060\,V resulting in an eletrical field of 8.3\,kV/cm.
	
	The whole device is contained in a 28.5$\times$38$\times$32.5\,cm$^3$ plexiglass cube that includes also the high-voltage power supply and complete data acquisition chain. The list mode data are saved on a memory card and are downloaded by wireless network every few days to a computer for offline analysis.
	
	The total material budget of the device, including the plexiglass box, is 2.63 g/cm$^2$, which corresponds to the projected range for 5\,MeV electrons. The device is thus insensitive to $\gamma$-rays from ambient radioactivity, which have energies $E_\gamma \leq$ 2.615 MeV. No lead shielding is used, and as a consequence in surface-based measurements the muon telescope is also sensitive to other cosmic rays or cosmic-ray induced particles, in particular high-energy electrons (roughly 10\% of the signal at the Earth's surface) and protons \cite{Olah17}.
	
	The signal threshold for each pad or field wire is fixed well above the electronic noise, at $\sim$400\,mV. 
	The data acquisition is triggered when at least two chambers fire in temporal coincidence ($\Delta t \leq$ 2\,\textmu s). Then, for each of the six chambers the firing pads and field wires are recorded, as well as the time of the event (with 400\,\textmu s granularity). 
	
	The nominal opening angle of the device is $\pm$56$^\circ$. However, already at an angle of $\theta =$40$^\circ$ the effective detection area is just one third the effective detection area for $\theta$ = 0$^\circ$ (vertical). Therefore, it is more practical to use several different orientations, if a full angular coverage is required. The REGARD muon telescope can be operated in any orientation.
	
	\begin{figure}[tb]
		\centering
		\includegraphics[width=\columnwidth]{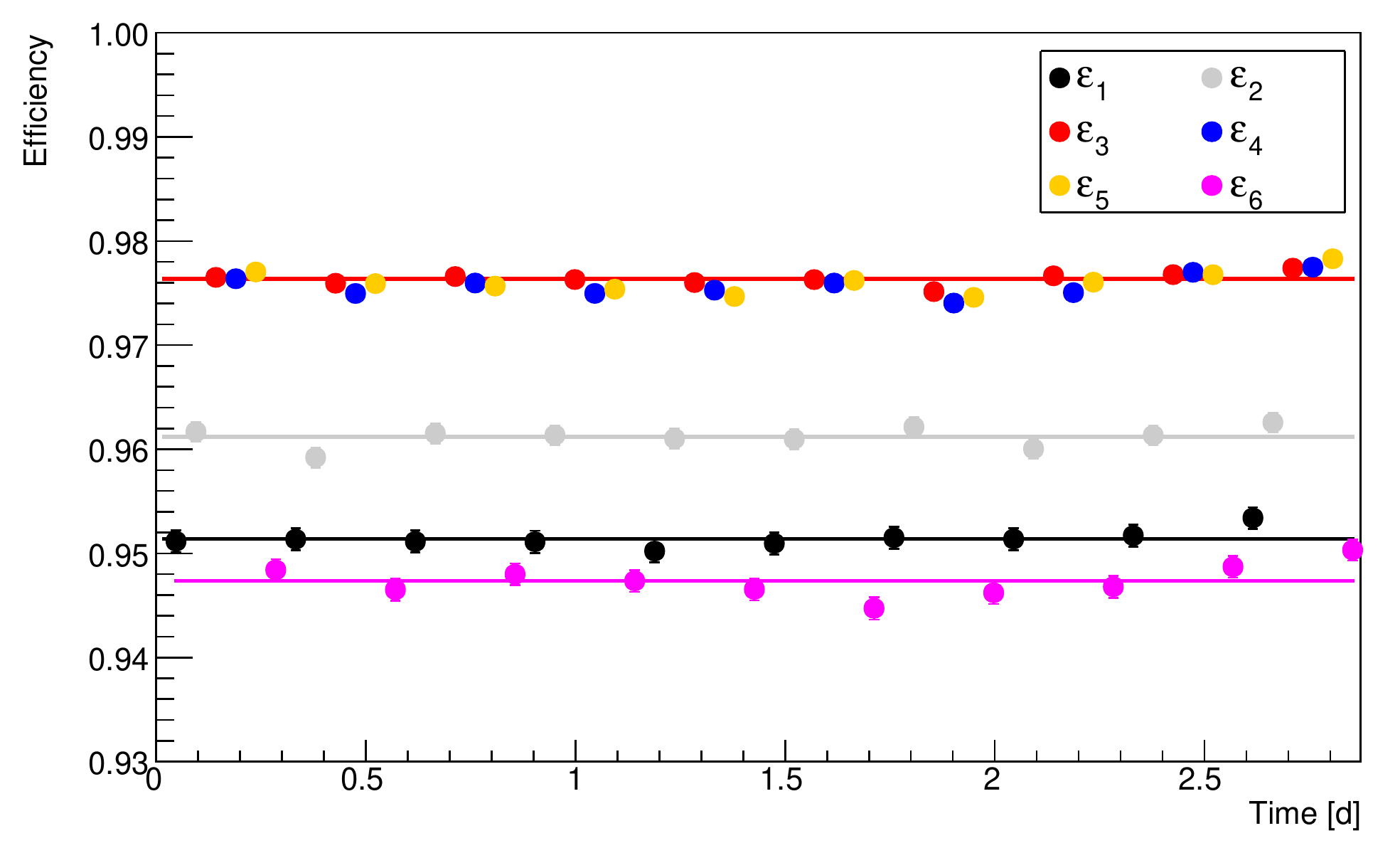}
		\caption{%
			Efficiency $\varepsilon_i$ of the individual chambers, with $\varepsilon_1$ being the top and $\varepsilon_6$ being the bottom chamber, as a function of time during a surface-based measurement. The statistical error bars are smaller than the symbols used. The data are fit with straight lines. For clarity, the lines for $\varepsilon_4$ and $\varepsilon_5$ are omitted. See text for details.}
		\label{fig:Efficiency}
	\end{figure}
	
	
	\subsection{Tests of the detector performance}
	
	As a first step, the detector was tested at the surface of the Earth, where due to the sizable muon and electron intensity a trigger rate of 17.24$\pm$ 0.12\,s$^{-1}$ is observed. 
	
	Out of the data recorded, in the offline analysis muon events were defined to be characterized by at least five of the six chambers firing in temporal ($\Delta t \leq$ 2\,\textmu s) and spatial (reconstructed track has a $\chi^2$ of less than 2) coincidence. Furthermore all reconstructed trackpoints have to be inside the detector volume. The individual muon detection efficiency $\varepsilon_i$ for each of the six chambers $i$ was determined by calculating a track without chamber $i$. $\varepsilon_i$ is then given by the number of tracks, where there is a hit channel on or near the intersection point of the calculated track in chamber $i$ divided by the total number of tracks (Fig.~\ref{fig:Efficiency}).
	
	Chambers 3, 4, and 5 have an efficiency of $\varepsilon_{3/4/5}\approx$ 0.97 (Fig.~\ref{fig:Efficiency}). The 3\% inefficiency is explained by spacers, which are holding the wires in position and therefore create dead zones for detection, and support pillars. The efficiency of the three other chambers is slightly lower, $\varepsilon_{1/2/6}\approx$ 0.95-0.96.

	The total efficiency  of the REGARD muon telescope is then given by:
	\begin{equation}
	\varepsilon = \prod\limits_{i = 1}^{6} \varepsilon_i + 
	\sum\limits_{i=1}^{6} \prod\limits_{j = 1}^{6}
	\left[ (1-\delta_{ij})\varepsilon_i + \delta_{ij}(1-\varepsilon_i)
	\right]
	\end{equation}
	where the usual Kronecker symbol $\delta_{ij}$ gives $\delta_{ij}$ = 1 for $i=j$ and 0 otherwise.
	
	\begin{figure}[tb]
		\centering
		\includegraphics[width=\columnwidth]{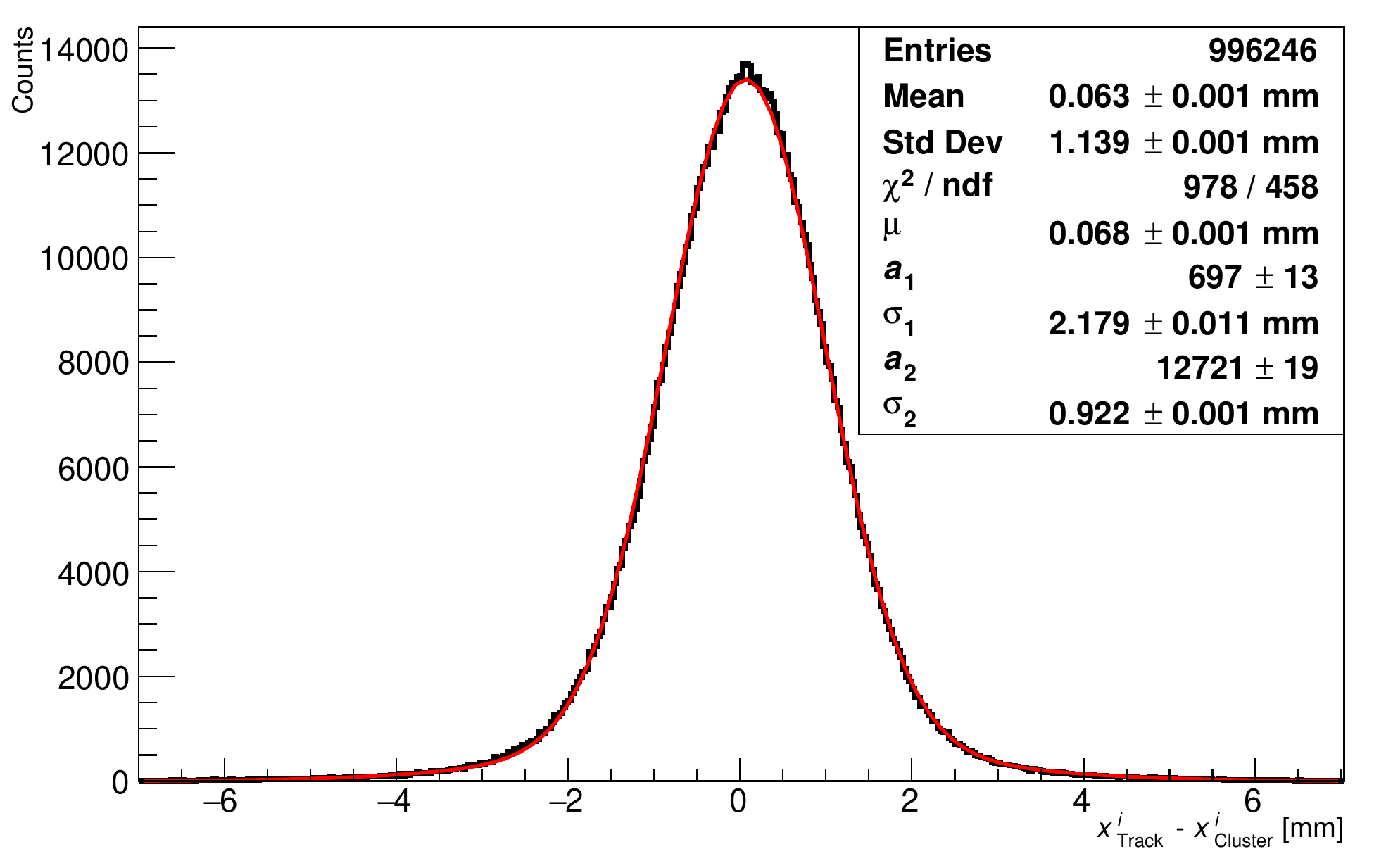}
		\caption{%
			Difference between the x coordinates of the calculated track point $x^i_{\rm Track}$ and of the cluster in the channel data $x^i_{\rm Cluster}$. The red curve is a fit of the sum of two Gaussian distributions $\sum_{i=1}^{2} a_i \exp [-(x-\mu)/2\sigma_i^2]$ with the same mean $\mu$ and different normalizations $a_i$ and widths $\sigma_i$. See text for details.}
		\label{fig:Alignment}
	\end{figure}
	
	\begin{figure*}[bt]
		\centering
		\includegraphics[width=\textwidth]{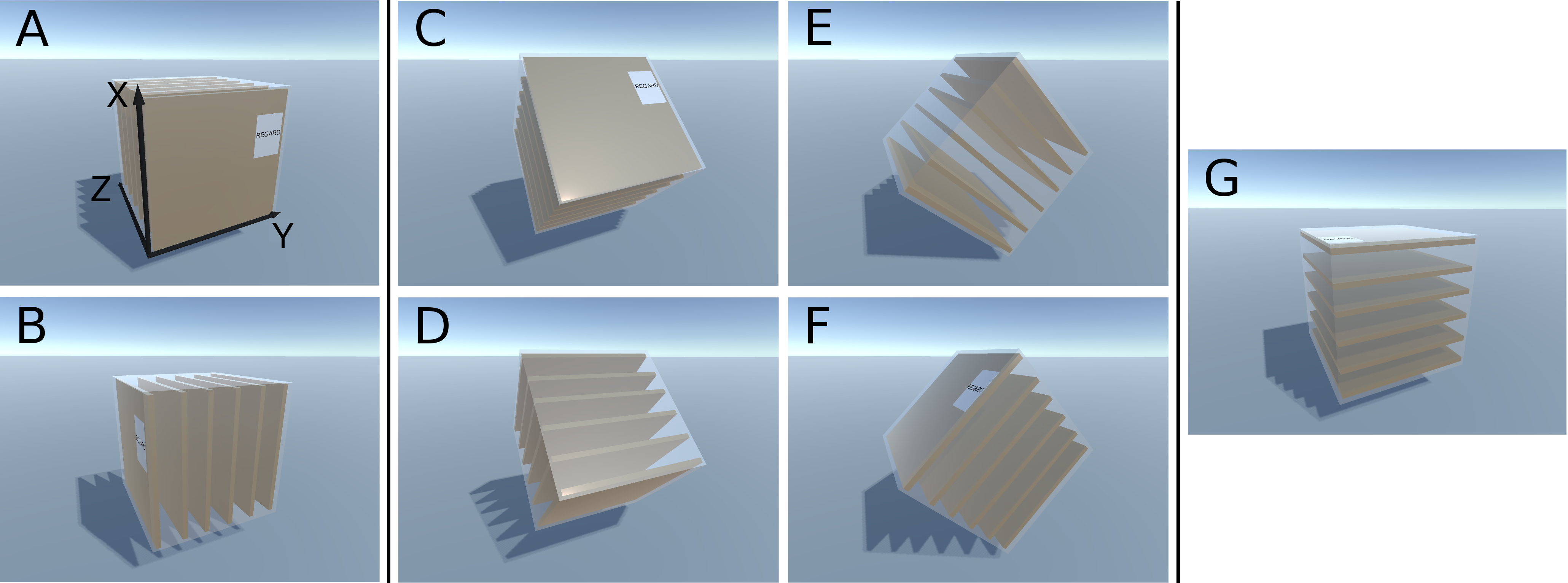}
		\caption{Sketch showing the seven different orientations of the telescope that were used, in turn, at each of the eight measurement positions: A and B show the vertical, C to F the 45\textdegree\ tilted and G the horizontal orientation. Panel A includes the coordinate system adopted in this work.}
		\label{fig:Orientations}
	\end{figure*}
	
	The surface-based data have subsequently been used to correct for misalignments of the chambers  (Fig.~\ref{fig:Alignment}). To this end, for each of the six chambers $i$ the difference between the fitted track position at this chamber height $y^i_{\rm Track}$ and the cluster in the field wire data $y^i_{\rm Cluster}$ is plotted. See Figure \ref{fig:Orientations} A for the definition of the $x$, $y$, and $z$ axes. The alignment data follow a Gaussian distribution, with some enhancement on the shoulders (Fig.~\ref{fig:Alignment}). 
	
	The mean of distribution $i$ is then used to correct the alignment of chamber $i$. Based on the above described procedure, the corrections are $\left|y^i_{\rm Track} - y^i_{\rm Cluster}\right|$ = 0.06 to 0.5\,mm for the field wires and $\left|x^i_{\rm Track} - x^i_{\rm Cluster}\right|$ = 0.016 to 0.2\,mm for the pads. 
	
	The alignment data (Fig.~\ref{fig:Alignment}) histogram the difference between two different estimates of the true position of a muon crossing detector plane $i$. Therefore, the fitted width $\sigma_{x,y}$ of the distribution gives an experimental upper limit on the spatial resolution of the telescope. The fits for the six detector planes give widths in the $\sigma_x$ =1.08-1.29\,mm and $\sigma_y$ = 0.97-1.23\,mm ranges for x and y axis, respectively. These values are consistent with earlier resolution data that had been obtained with a pencil beam of minimum ionizing particles incident on another copy of the same device \cite{Varga13}.
	
	Using the experimentally determined $\sigma_{x,y}$ values and the 17.5\,cm height of the muon telescope, an expected angular resolution of $\leq$7.4\,mrad, or $\leq$0.85$^\circ$, is found. This value gives a lower limit for the size of the angular bins used when histogramming the data. 
	
	\subsection{Measurement procedure}
		
	In order to obtain data with complete angular coverage and satisfactory statistics, also at angles that are disfavored by the muon angular distribution, at each of the eight positions (see Sec.~\ref{sec:Felsenkeller}) studied, seven different orientations of the muon telescope have been used (Fig.~\ref{fig:Orientations}) in sequence.
	
	For the underground measurements, running times of 3-4 days were used for each orientation, while on the surface a measurement time of half a day per orientation proved sufficient. The observed trigger rates were in the range of 4.7 to 5.2\,s$^{-1}$  for all orientations in tunnels~VIII and IX, caused by the radioactivity of the surrounding rock. For tunnel~4, where the surrounding rock is shielded by concrete, the trigger rates vary from 0.8 to 2.0\,s$^{-1}$.

	\section{Data analysis and results}
	\label{sec:Analysis}
	
	\subsection{Data sorting}
	\label{subsec:Sorting}
	
	As a first step in the offline analysis, the above described criterion of five chambers firing was applied. Subsequently, for each preliminarily identified muon event, the tracking algorithm \cite{Olah16phd} was used to calculate the most likely muon track. The resulting rate of detected muon tracks underground ranged from 0.020 $\pm$ 0.006\,s$^{-1}$ for orientation~A to 0.1009 $\pm$ 0.0013\,s$^{-1}$ for orientation~G (see Fig.~\ref{fig:Orientations} ). 
	
	For each track, subsequently the coordinates in the telescope's proper coordinate system were converted to zenith and azimuthal angles $\theta$ and $\phi$, taking into account the orientation of the muon telescope during that particular run.
	
	\subsection{Determination of the muon intensity}
	\label{subsec:Muonintensity}
	
	An angular grid of step size 10$^\circ$ and 20$^\circ$ for the zenith and azimuthal angles $\theta$ and $\phi$, respectively, was then used for all further analysis steps. 
	Subsequently, separately for each detector orientation A-G the effective detection area of the detector was determined for each track.

	After that, the muon track data from this site and orientation were filled into a two-dimensional histogram, which was then rescaled bin per bin by the measured efficiencies $\varepsilon_i$ for each chamber $i$, the solid angle and the live time of the measurement, giving a histogram for the muon intensity.
	
\begin{figure}[h!!]
		\centering
		\includegraphics[width=\columnwidth]{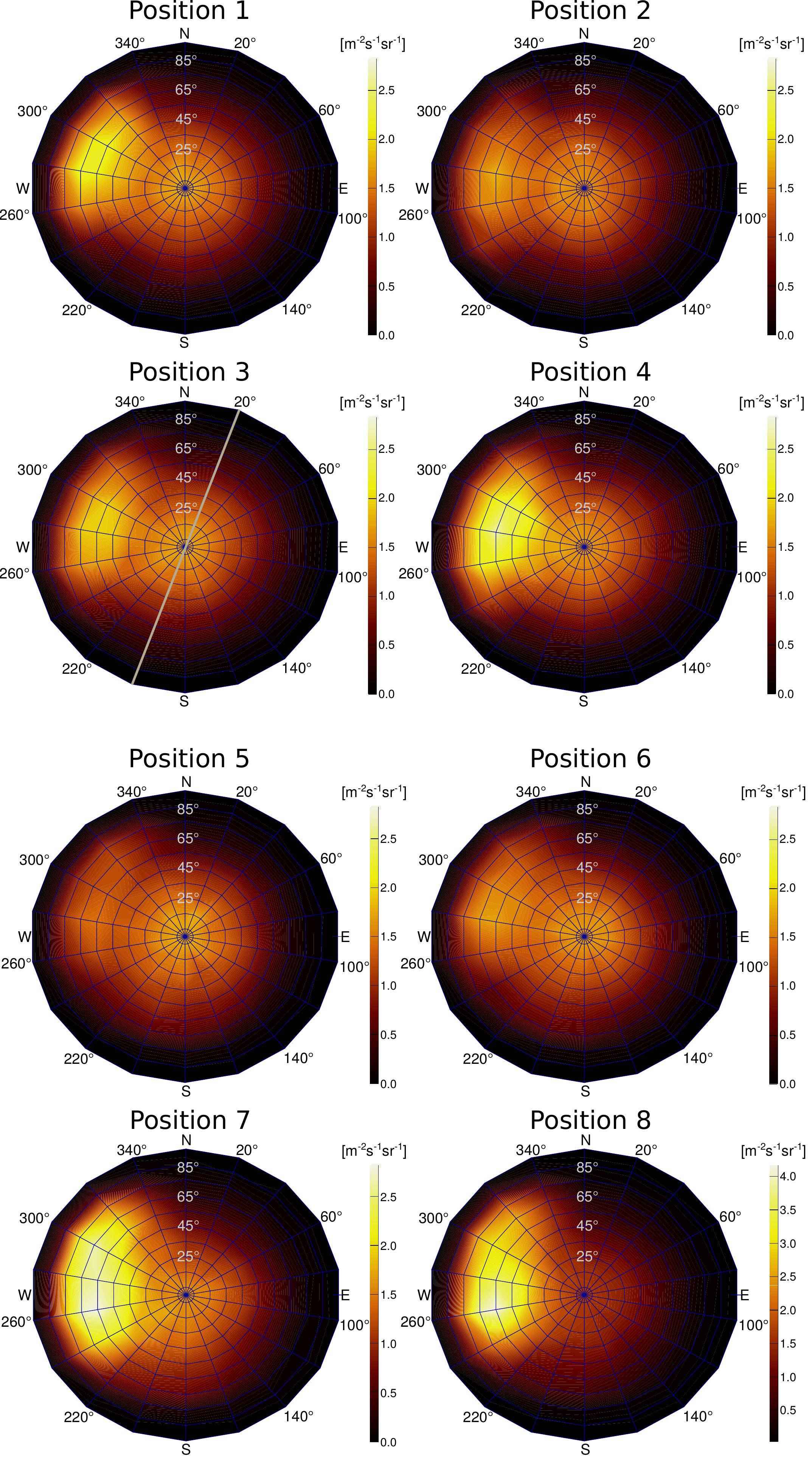}
		\caption{Muon intensity at positions 1-4 (tunnels VIII and IX) and 5-8 (tunnel~IV) of Felsenkeller. Note the different color scale for position 8, which is closest to the tunnel entrance. All muon intensities are in $\left[\text{m}^{-2}\text{s}^{-1}\text{sr}^{-1}\right]$. The grey line in position~3  marks the values taken for Fig.~\ref{fig:Angularfit}.}
		\label{fig:ExpMuonintensity2DMaps}
	\end{figure}

	The seven separate histograms A-G were then checked for consistency bin by bin, and the relative deviations were found to follow the expected normal distribution. After that the data from the separate histograms were used for a weighted average, taking into account their individual statistical errors, to develop the complete intensity map for this position (Fig.~\ref{fig:ExpMuonintensity2DMaps}).  
	
	The statistical errors of the muon intensity map are in the order of 3\% for small zenith angles and $\leq$12\% for $\theta \leq 75^\circ$. For $\theta = 85^\circ$, the statistical uncertainties are usually $\sim$20\%, and up to 40\% for single bins due to the low muon intensity.
	
	\subsection{Systematic uncertainties in the muon rate}
	\label{subsec:Uncertainties}
	
	Since the telescope cannot distinguish between upgoing and downgoing muon tracks, these tracks contain a small number of upgoing muon events. The intensity of neutrino-induced muons is of the order of $10^{-8}\,\text{m}^{-2}\text{s}^{-1}\text{sr}^{-1}$ \cite{Grieder01}, negligible at the 45\,m depth of Felsenkeller.
	
	In cases where more than one charged particle passes through the detector within the data acquisition time window for event building (400\,\textmu s), one of the two events will be rejected by the tracking algorithm.  Based on the above mentioned track rate and time window, it is estimated that this effect leads to a decrease of the calculated muon intensity by less than 1\%, negligible for the present purposes. 

Possible false muon signals may in principle stem from electrons induced by muons in the rock that exceed the muon telescope's effective energy threshold of $E_e \geq$ 5\,MeV for electrons \cite{Olah17}. However, the simulations (see below, Sec.~\ref{subsec:MonteCarlo}) showed they usually coincide with a surviving muon and are therefore rejected by the tracking algorithm.

	This is different for muon-induced electrons traveling roughly in the original direction of the muon that induced them. If they have sufficient energy after passage through the rock cover and the air inside the tunnel before hitting the muon detector, they may lead to a false muon signal. However,  that to good approximation muon-induced electrons always coincide with a surviving original muon. The REGARD tracking algorithm only accepts one track per event, so in such a case it would count only one muon and discard the electron track.

	\subsection{Interpretation of the muon intensity data}
	\label{subsec:Interpretation}
	
	It is clear from the muon intensity maps (Fig.~\ref{fig:ExpMuonintensity2DMaps}) that instead of the expected maximum at $\theta$ = 0$^\circ$, there is a wide maximum in northwesterly direction at a zenith angle of $\theta_{\rm max}$ = 55-65$^\circ$. Seen from the measurement positions, the cliff of the Weißeritz valley is located in this approximate direction (Fig.~\ref{fig:Terrain}). Still, there is a secondary maximum at $\theta$ = 0$^\circ$ for all eight positions studied.
	
	The direction and value of the maximum intensity $I_{\rm max}$ and also the integrated intensity $J$, which is calculated by integrating I over all angles, for each position are listed in Table~\ref{Table:Muonintensity}.
	
	The intensity maps show a homogeneous overall pattern for all positions studied, with one notable exception: The maximum in the direction of the cliff is much more pronounced for positions 3, 4, and 7, which show $I_{\rm max}$ values between (1.96$\pm$0.05) and (2.82$\pm$0.05) m$^{-2}$s$^{-1}$sr$^{-1}$, than for the three positions 2, 5, and 6 deepest inside the tunnels: Positions 2 and 6 show $I_{\rm max}$ = (1.80$\pm$0.04) and (1.71$\pm$0.04) m$^{-2}$s$^{-1}$sr$^{-1}$, respectively. For position~5, the muon intensity in the direction of the cliff is just (1.46$\pm$0.04) m$^{-2}$s$^{-1}$sr$^{-1}$, lower than the vertical flux at that position (Table \ref{Table:Muonintensity})
	
	The highest intensity, $I_{\rm max}$ = (4.17$\pm$0.07) \linebreak m$^{-2}$s$^{-1}$sr$^{-1}$, is observed at position 8, which is the measurement position closest to the tunnel entrance. 
	
	In the directions away from this westerly-northwesterly intensity maximum, i.e. North, East, and South, an intensity in the 0-2 m$^{-2}$s$^{-1}$sr$^{-1}$ range is observed. This is consistent with the known terrain shape (Fig.~\ref{fig:Terrain}), an essentially flat area extending over at least 100\,m in these three directions beyond the measurement points studied. 
	
	\begin{table}[t]
		\begin{tabular}{|r|r|r|d{4}|d{4}|}
			\hline 
			Position & $\theta_{\rm max}$  & $\phi_{\rm max}$ & \multicolumn{1}{c|}{$I_{\rm max}$} & \multicolumn{1}{c|}{$J$} \\ 
			& [$^\circ$] & [$^\circ$] & \multicolumn{1}{c|}{[m$^{-2}$s$^{-1}$sr$^{-1}$]} & \multicolumn{1}{c|}{[m$^{-2}$s$^{-1}$]} \\ \hline \hline
			1 & 65$\pm$5 & 280$\pm$10 & 2.26(5) & 5.0(4) \\
			2 & 55$\pm$5 & 280$\pm$10 & 1.80(4) & 4.6(5) \\
			3 & 55$\pm$5 & 300$\pm$10 & 1.96(5) & 4.9(4) \\
			4 & 55$\pm$5 & 280$\pm$10 & 2.66(6) & 5.4(4) \\ \hline \hline
			5 &  0$\pm$5 & --- & 1.73(2) & 4.6(4) \\
			6 & 55$\pm$5 & 280$\pm$10 & 1.71(4) & 4.7(4) \\
			7 & 55$\pm$5 & 260$\pm$10 & 2.82(5) & 6.0(4) \\
			8 & 55$\pm$5 & 260$\pm$10 & 4.17(7) & 7.0(4) \\ \hline
		\end{tabular}
		\caption{%
			\label{Table:Muonintensity}%
			For each measurement position, the direction ($\theta_{\rm max}$,$\phi_{\rm max}$) and value $I_{\rm max}$ of the highest muon intensity and the integrated muon intensity $J$, integrated over all angles, are listed. For $I_{\rm max}$ and $J$, only statistical errors are shown.} 
	\end{table}
	
	These differences in $I_{\rm max}$, by up to a factor of 2.4 for tunnel IV and 1.5 for tunnel VIII, are somewhat attenuated when instead the integrated intensity is considered: The largest relative $J$ differences are by a factor of 1.5 for tunnel IV and 1.2 for tunnel VIII, mainly due to the largely unchanged intensity in all directions except for the "cliff" maximum. 
	
	
	After the present measurement campaign was completed, at positions 3 and 4 two measurement bunkers for the new underground accelerator laboratory have been erected \cite{Bemmerer18-SNC}. These two positions show an integrated muon intensity $J$ = 4.9-5.4 m$^{-2}$s$^{-1}$, 4-15\% higher than position 6 (tunnel IV, bunker MK1) but 10-18\% lower than position 7 (tunnel IV, bunker MK2). In order to compensate the slight disadvantage of the two new bunkers with respect to the optimal existing bunker MK1, it seems advisable to install a muon veto there, especially in the direction of the intensity maximum.
	
	
	\begin{figure}[h]
		\centering
		\includegraphics[width=\columnwidth]{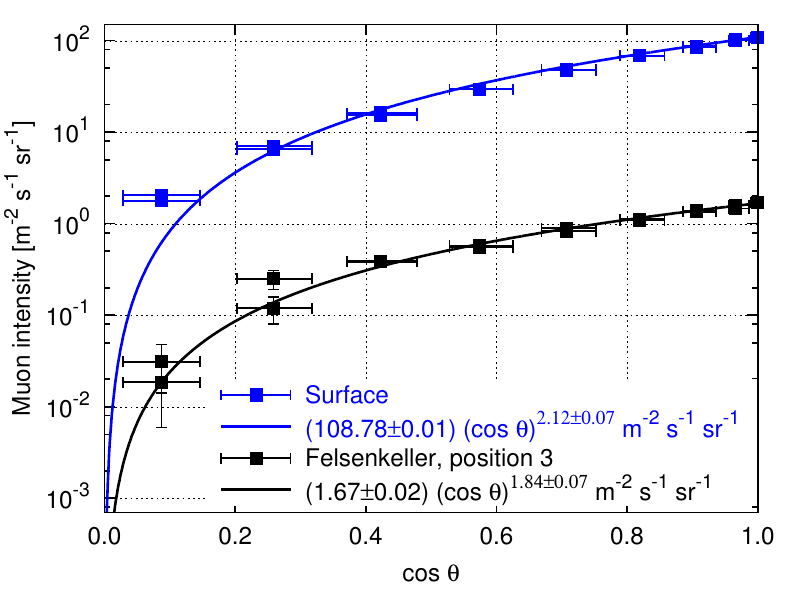}
		\caption{Fit of one underground (position 3) and one surface measurement to the $|\cos\theta|^n$ law. The $\phi$ direction selected is marked with a thin grey line in Fig.~\ref{fig:ExpMuonintensity2DMaps}. Each data point represents a 10$^\circ$ wide bin in $\theta$, either on the right side ($\phi$=20$^\circ$) or on the left side ($\phi$=200$^\circ$) of the distribution. The horizontal error bar has been chosen such that 68\% of the muon tracks in the bin are inside the area covered by the error bar. See text for details.}
		\label{fig:Angularfit}
	\end{figure}
	
	\subsection{Determination of the zenith angle distribution}
	\label{subsec:Zenith}
	
	In order to verify whether the zenith angle dependence follows a $|\cos(\theta)|^2$ angular distribution, in the position 3 intensity map, a 20$^\circ$ wide slice in $\phi$ is studied. The angular cut is selected at $\phi$ = (20$\pm$10)$^\circ$ (right half) and (200$\pm$10)$^\circ$ (left half), in order to take advantage of the flat surface above and to avoid the intensity maximum that is due to the cliff (sec.~\ref{subsec:Interpretation}). The center of the $\phi$ range selected is shown by a thin grey line in Figure \ref{fig:ExpMuonintensity2DMaps} at position 3. 
	
	When fitting these data, an exponent of $n$ = 1.84$\pm$0.07 is found (Fig.~\ref{fig:Angularfit}).  This value is slightly below the previous value of $n=2$ from a tunnel under the Mont Blanc at 140\,m.w.e. depth \cite{Castagnoli65}. At the Earth's surface, here $n$ = 2.12 $\pm$0.07 is found, consistent with previous work \cite{Crookes72}. However, the present detector is also sensitive to the soft component of cosmic rays. This might skew the observed angular dependence for the surface-based data but not for the underground tunnels, where the component is removed.
	
	It should be noted that the limited statistics required angular bins that are significantly larger than what is in principle possible with the REGARD muon telescope. A finer angular binning would have required higher measurement times, which for practical reasons were not attempted here.

	\section{Calculation of the predicted muon intensity}
	\label{sec:Calculation}
	
	In order to verify whether the terrain description is complete, as a next step the predicted muon intensity was calculated. For the calculation, position 3 was selected, the place for one of the measurement bunkers of the future accelerator laboratory \cite{Bemmerer18-SNC}. 
	
	\subsection{Determination of the rock thickness and properties}
	\label{subsec:RockProperties}
	
	The terrain features (Fig.~\ref{fig:Terrain}) were taken from the official geodata "Digitales Geländemodell 1" DGM1 with a grid size of 1.0\,m. These data have been obtained by aircraft-based laser scanning of the terrain. For the calculation, the last echo point was used, meaning that previous echos due to vegetation were discarded. In cases where there was only one echo point, the only echo point was used. The grid points have a location uncertainty of 0.2\,m and an elevation uncertainty of 0.15\,m. The DGM1 data are provided digitally by Staatsbetrieb Geobasisinformation und Vermessung Sachsen, Dresden/Germany, in the ETRS89\_UTM (European Terrestrial Reference System \linebreak 1989, UTM zone 33) position reference system and \linebreak DHHN2016 (Deutsches Haupthöhennetz 2016) elevation reference system. 
	
	The insides of tunnels VIII and IX were scanned by three-dimensional georeferenced laser scanning, performed by Ingenieurbüro Leibiger, Kesselsdorf/Sachsen, in the position reference system RD/83 (a transformation of the 42/83 reference system to the Rauenberg datum that was applied for Eastern Germany) and the elevation reference system DHHN92. The tunnel data were provided with a precision of $\leq$1\,cm. The difference between the two different elevation reference systems DHHN2016 and DHHN92 used here amounts to just [0cm;+3cm] in Saxony and was neglected for the present purposes.
	
	For Fig.~\ref{fig:Terrain} and the calculations and simulations performed in the present work, the DGM1 data were transformed from the original ETRS89\_UTM system to RD/83 by the tool provided by Staatsbetrieb Geobasisinformation und Vermessung Sachsen, with a resultant additional uncertainty of a few cm, negligible for the present purposes.

	\begin{figure*}[bt]
		\centering
		\includegraphics[width=\textwidth]{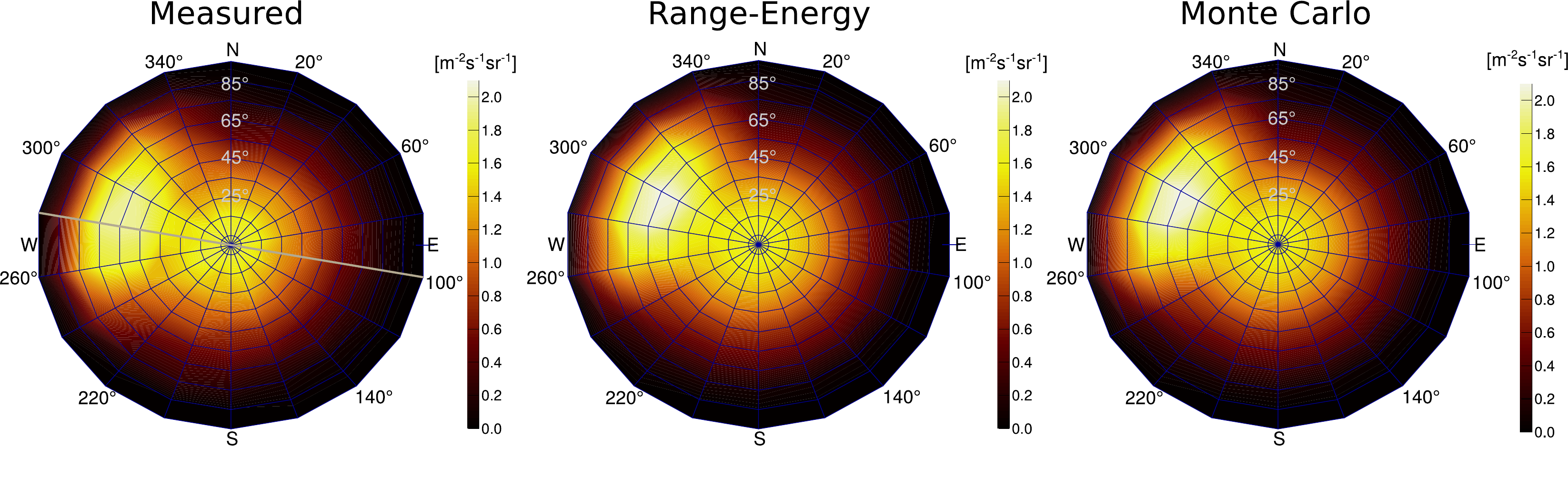}
		\caption{Muon intensity map for position 3 (same as in Fig.~\ref{fig:ExpMuonintensity2DMaps}, but with slightly modified color range): From left to right: Measured, calculated from the range-energy relation (sec.~\ref{subsec:RangeEnergy}), and Monte Carlo simulated (sec.~\ref{subsec:MonteCarlo}) muon intensity. The grey line marks the cut used for Fig.~\ref{fig:CutExpCalcMonteCarlo} below.}
		\label{fig:ExpCalcMonteCarlo}
	\end{figure*}
	
	\subsection{Calculation of the muon intensity based on the range-energy relation}
	\label{subsec:RangeEnergy}
	
	The usual Gaisser parameterization of the muon intensity had been developed for deep-underground scenarios \cite{Gaisser90}, where high-energy muons play a decisive role. Thus, the original formula only applies for $E_\mu >$ 100 GeV/$\cos\theta$.
	Recently, Tang and co-workers \cite{Tang06} proposed some modifications of the Gaisser parameterization for scenarios where low-energy muons play a role, extending the range of validity to 1\,GeV/$\cos\theta$ and below. Here, the approach by Ref. \cite{Tang06} is followed and called Gaisser-Tang.
	
	In order to model the energy loss by muons in the rock overburden, Gaisser's range-energy relation \cite{Gaisser90,Grieder01} is adopted here:
	\begin{equation} \label{eq:RangeEnergy}
	E = \left(\frac{a}{b}\right) \left[\exp(b R_\mu)-1\right]
	\end{equation} 
	with $a = 0.217\,\text{GeV\,m.w.e.}^{-1}$, $b = 4.5 \cdot 10^{-4}\,\text{m.w.e.}^{-1}$ and $R_\mu$ the muon range in m.w.e. For the energy loss, no special low-energy modifications are necessary.
	
	Based on the range-energy relation (\ref{eq:RangeEnergy}), for a given slant depth $x$ a cutoff energy $E_{\rm cutoff}(x)$ has been calculated. Muons which have lower energy than $E_{\rm cutoff}(x)$ are assumed to be absorbed in the rock, and secondary particles are neglected for the present purposes.
	
	Since the Gaisser-Tang parameterization depends on the zenith angle $\theta$, the calculation was done with a total of 60 azimuthal and 30 zenith bins. For each bin, the slant depth $x$ was calculated from the geodetic data on the tunnels and the rock overburden and the known rock properties (sec.~\ref{subsec:RockProperties}, Fig.~\ref{fig:Terrain}). Depending on the direction for position~3, $x$ varies between 60\,m.w.e. in the direction of the tunnel entrance and 7000\,m.w.e. for large zenith angles $\theta$. 
	
	Subsequently, for each angular bin, the muon energy spectrum was integrated starting at the cutoff energy \linebreak $E_{\rm cutoff}(x)$, and the resulting muon intensity was retained. The effective energy threshold of the muon telescope is just 30\,MeV \cite{Groom01} and was neglected here. Subsequently, the histogram was rebinned to the rougher binning used to present the experimental data.
	
	The resulting calculated intensity is shown in Fig.~\ref{fig:ExpCalcMonteCarlo}, central panel. In addition, a 20$^\circ$ wide $\phi$ cut of the same results around $\phi = \left[100-120\right]^\circ$/$\left[280-300\right]^\circ$ is shown in Fig.~\ref{fig:CutExpCalcMonteCarlo}. The cut angle was chosen to include the "cliff" flux maximum (sec. \ref{subsec:Interpretation}), i.e. the most complicated part of the terrain to reproduce.

		
	\subsection{Monte Carlo simulation}
	\label{subsec:MonteCarlo}
	
	As an independent second approach to determine the predicted muon flux, a Monte Carlo simulation of the muons passing through the rock was performed. 
	
	Since the frequently used Monte Carlo codes MUSUN and MUSIC \cite{Kudryavtsev09} are mainly suitable for large depths, here instead Geant4 (version 10.4) \cite{Agostinelli03} with the physics list "Shielding2.1" and the option "EMZ" was used. 
	
	As in sec.~\ref{subsec:RangeEnergy} above, the Gaisser-Tang muon intensity parameterization \cite{Tang06} was used, and again the slant depth $x$ was calculated from the geodetic data and the known rock properties for a total of of 60 azimuthal and 30 zenith bins. For each bin, a total of $2\times 10^6$ muons were propagated through a 10\,m$\times$10\,m wide column of rock, with a height given by the slant depth. 
	
	For the slant depths relevant in this work, the effect of multiple scattering of muons \cite{Olah19} is of the order of 10-12\,mrad \cite{Olah18-privcomm}, much 
	smaller than the bin sizes chosen here. The simulation shows just $10^{-6}$ muons and other particles lost to the sides of the 10\,m wide column for each muon detected in the telescope, negligible for the present purposes. 
			
\begin{figure}
		\centering
		\includegraphics[width=\columnwidth]{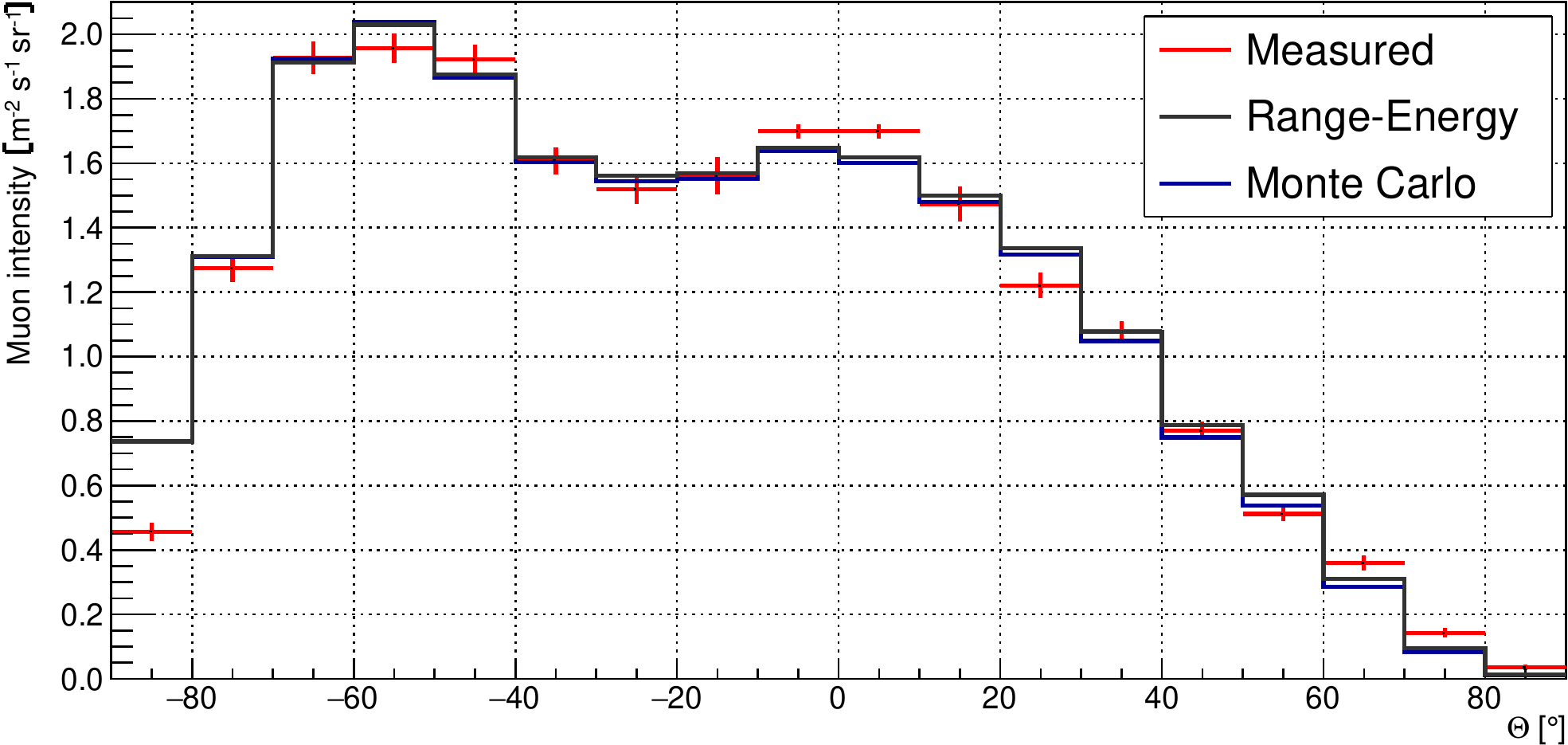}
		\caption{%
	\label{fig:CutExpCalcMonteCarlo}
Muon intensity at position~3 for $\phi = 280$\textdegree\ for negative $\theta$ and $\phi=100$\textdegree\ for positive $\theta$: Measured intensity (red), calculated based on the range-energy relation (black), Monte Carlo simulation (blue). }
	\end{figure}

	The resulting intensity from the Monte Carlo simulation is shown in Fig.~\ref{fig:ExpCalcMonteCarlo}, right panel, and a 20$^\circ$ wide $\phi$ cut in Fig.~\ref{fig:CutExpCalcMonteCarlo}.
	
	\section{Discussion}
	\label{sec:Discussion}

	\subsection{Comparison of the data with the predictions}
	
	At all locations studied, the angular distribution of the muon intensity and also the angle-integrated intensity are similar. 
	The angular data show two maxima: In addition to the expected maximum at zenith angle $\theta$ = 0$^\circ$, an even stronger second maximum is found at $\theta$ = 55$^\circ$, $\phi$ = 280$^\circ$, in the direction of a cliff in the rock overburden. 
	
	Both the prediction based on the muon range-energy relation and the Geant4 Monte Carlo simulations reproduce both the angular dependence and the absolute flux. A small deviation is found in the western direction ($\theta$ = -85$^\circ$ in Fig.~\ref{fig:CutExpCalcMonteCarlo}, where the measured intensity is lower due to buildings and the opposite rock wall of the river valley, which were not included when calculating the rock thickness. 
	
	The angle-integrated intensities of the different methods show excellent agreement with the measurement, as well: The range-energy calculation gives $J_{\rm RE}$ = 5.0\,m$^{-2}$s$^{-1}$, the simulation  $J_{\rm MC}$ = 4.9\,m$^{-2}$s$^{-1}$, and the experiment  $J_{\rm exp}$ = 4.9(4)\,m$^{-2}$s$^{-1}$. It can be concluded the rock overburden is correctly described by the geodetic data and the parameters $a$ and $b$ for the range-energy relation eq. (\ref{eq:RangeEnergy}) are appropriate for the present hornblende monzonite rock.

	\begin{figure}
		\centering
		\includegraphics[width=\columnwidth]{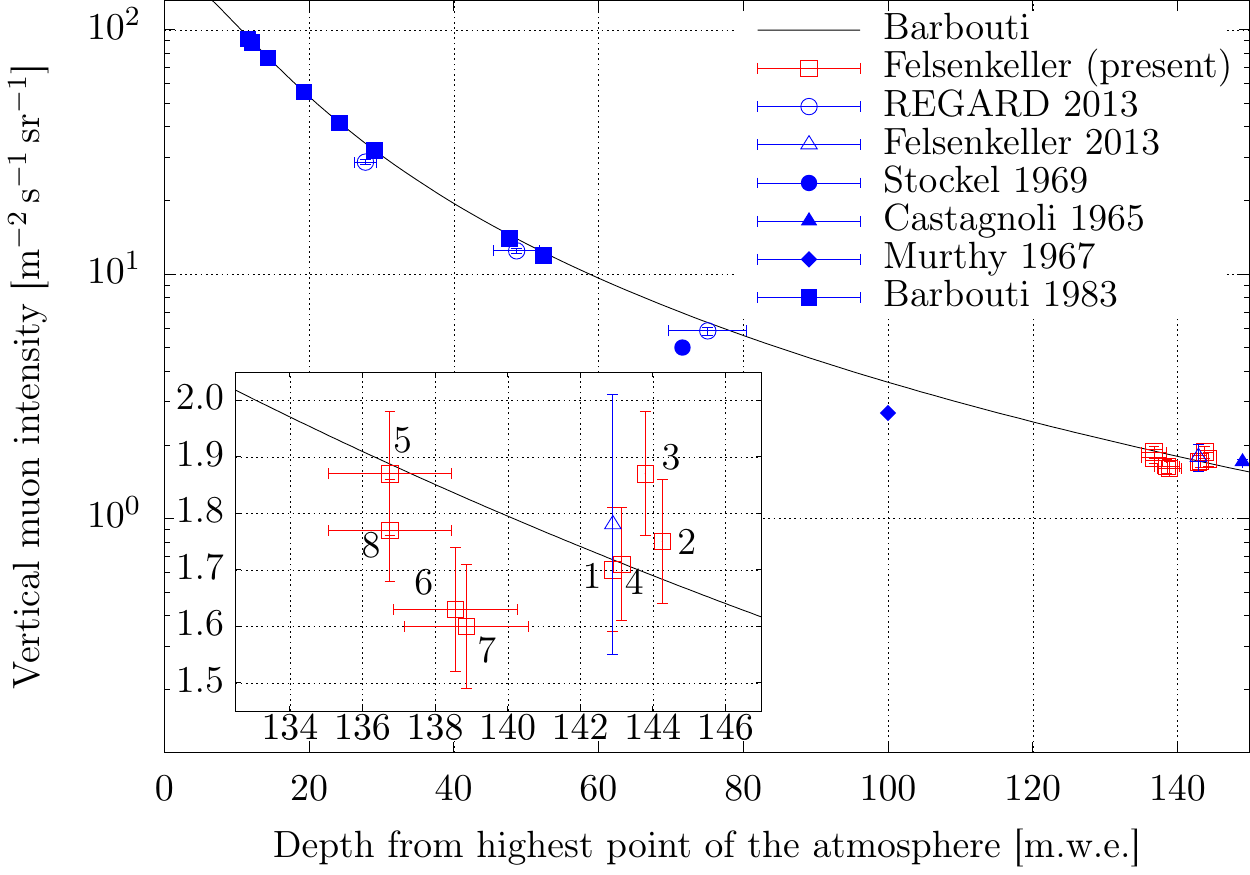}
		\caption{Vertical muon intensity (4$^\circ$ opening angle) as a function of depth, measured from the top of the atmosphere. The filled blue data points \cite{Stockel69,Castagnoli65,Murthy67,Barbouti83} are taken from tables in Ref. \cite{Grieder01}. In addition, previous REGARD data from Budapest/Hungary \cite{Olah16phd} (open blue circles) and the present data (red open squares) are shown. For the previous measurement in Felsenkeller by REGARD \cite{Olah16} (open blue triangle) the depth was updated to the new density measurement. The curve is the parameterization suggested by Barbouti {\it et al.} \cite{Barbouti83}. The inset shows a detailed view of the nine measurements in Felsenkeller in linear scale. Systematic errors on the depth due to the uncertainty of the rock density are not included in the horizontal error bars.}
		\label{fig:mwe}
	\end{figure}
	
	\subsection{Comparison of the vertical muon intensity with previous data}
	
	The vertical muon intensity measured in the present work (taken from a 4$^\circ$ wide bin around $\theta$ = 0$^\circ$, Fig.~\ref{fig:mwe}) is in excellent agreement with the previous parameterization of the muon intensity given by Barbouti {\it et al.} \cite{Barbouti83}. This lends further support to the conclusion that the rock overburden has been calculated correctly from the geodetic data and the rock properties. 
	
	The small variations, by up to 7 m.w.e., in rock overburden between the eight positions studied here do not lead to an appreciable variation in the vertical intensity within the present error bars (Fig.~\ref{fig:mwe}). 
	
	At position 8, the underground position with the highest angle-integrated intensity studied here (Table \ref{Table:Muonintensity}), both the rock overburden and the vertical intensity do not differ much from the other positions (Fig.~\ref{fig:mwe}, inset). 	
	The high integrated intensity observed at position 8 therefore has to be explained by the high flux of muons infalling from the cliff direction at $\theta$ $\approx$ 55$^\circ$ in this direction (Table~\ref{Table:Muonintensity}). 
	
	\subsection{Effects of variations of the muon flux with time on the present data}
	
	It is well known that the muon flux at the surface of the Earth varies over time, due to variations in the atmospheric temperature \cite{Grieder01}. The temperature variations affect the density, which, in turn affects the conversion from primary cosmic rays to muons \cite{Grieder01}. These effects usually have a period of one year, opposite signs for the Northern and Southern hemispheres, respectively, and amount to $\pm$2\% at the Earth's surface \cite{Mendonca16-ApJ}. In addition, long-term surface-based data from the Nagoya muon detector (35$^\circ$ latitude, lower than the present 51$^\circ$) show an overlaid variation that follows the 11-year solar cycle and has an amplitude of up to $\pm$3\% \cite{Mendonca17-IAU}, i.e. 1.5 times larger than the annual temperature variations. 

At the underground depth of Felsenkeller, 140 m.w.e., an average muon energy of 17\,GeV is expected. For this muon energy, there is an expected temperature coefficient \cite{Abrahao17-JCAP} of 
\begin{equation}
\alpha_{T} \equiv \frac{\Delta I / I}{\Delta T/T} \approx 0.2
\end{equation}
with $I$ and $\Delta I$ the average muon intensity and its variation, and $T$ and $\Delta T$ the average atmospheric temperature and its variation. Using the ground-based weather data at Dresden, within one year $\Delta T/T \approx$ 12 K/286 K = 0.04 is found. This leads to an expected $\Delta I/I \approx$ 0.2$\times$0.04=0.8\%. This value is similar to annual muon flux variations found experimentally in laboratories in the same depth range as Felsenkeller: The Double CHOOZ near detector at 120 m.w.e. (but with an energy threshold of 22\,GeV, higher than here) reported about $\pm$1\% fluctuations, and three laboratories at 34-250 m.w.e. showed $\pm$0.5\% fluctuations \cite{Sagisaka86-NCC}. 

For 17\,GeV muons, the expected variation due to variations $\Delta p$ in atmospheric pressure $p$ \cite{Sagisaka86-NCC} is much smaller, 
\begin{equation}
\beta_{p} \equiv \frac{\Delta I / I}{\Delta p/p} \approx 0.02
\end{equation}
%
so that pressure effects are negligible for the present purposes.

The present campaign did not run long enough to experimentally constrain the muon flux variation at Felsenkeller, therefore these effects were included in the error budget. From the above derived annual temperature effect of $\pm$0.8\% and an estimated additional 1.2\%, i.e. 1.5 times as much, due to the solar cycle \cite{Mendonca17-IAU}, a total effect of $\pm$2\% is found. This value is adopted as 2\% systematic error and included in the error budget. 

This uncertainty will, however, not affect the planned muon veto for nuclear astrophysics experiments. There, the veto trigger rate will also be recorded, allowing to correct for muon intensity variations. It is expected that as a byproduct of these experiments, long-term data on muon flux variations in Felsenkeller will become available in the future.
	
	\section{Conclusions, summary, and outlook}
	\label{sec:Summary}
	
	The REGARD muon telescope has been used to measure the muon intensity at eight different positions in the tunnels of the Felsenkeller shallow-underground site in Germany and at one position on the Earth's surface. 	
	Inside the  tunnels, the muon intensity is found to be rather homogenous and shows two maxima: one in vertical direction, the other one towards the rock cliff. 
	
	The dependence of the muon intensity on the zenith angle was examined both underground and overground. The intensity was found to follow the dependence $I\left(\theta\right) \propto \cos^n \theta $, with $n = 1.84 \pm 0.07$ underground and $n = 2.12 \pm 0.07$ on the Earth's surface.
	Given the small size of the muon telescope and the limited measurement time, the expected magnitude of muon intensity variations over time is smaller than the statistical error bars. No attempt was made to study such variations. 
	
	The present muon intensity data were matched both in the absolute values and in their angular pattern, first by calculations based on the range-energy relation of muons, and second by a Geant4 simulation. 	
	
	The effective vertical rock overburden of Felsenkeller tunnel~IV, site of a low-radioactivity counting facility established in 1982 \cite{Helbig84-Isotopenpraxis} and enlarged in 1995 \cite{Niese98} was found to be 138\,m.w.e., both by the terrain properties and by the observed vertical muon intensity. The integrated muon intensity is 30-40 lower than at the Earth's surface (190 m$^{-2}$s$^{-1}$ \cite{Grieder01}).
	
	For tunnels~VIII and IX, where a new low-background accelerator and $\gamma$-counting facility are being commissioned \cite{Bemmerer18-SNC}, a slightly higher rock overburden of 143\,m.w.e. was found. Again, the depth determinations were consistent both from the terrain features and from the muon intensity. In those tunnels, the integrated muon intensity is 35-40 times lower than at the surface. 
	
	The knowledge on the muon intensity and angular distribution will be instrumental in studies of muon-induced neutrons \cite{Grieger19-inprep} and in properly positioning muon veto detectors for the planned accelerator-based experiments in tunnels~VIII and IX. 
	
	\subsection*{Acknowledgments}
	
	The authors are indebted to Prof. Vitaly Kudryavtsev, University of Sheffield, for helpful suggestions on setting up the muon intensity calculations and simulations, to Dres. Gerg\H{o} Hamar, L\'aszl\'o Ol\'ah and Dezs\H{o} Varga, MTA Wigner Budapest, for their comments on the present manuscript, to Dr. Axel Renno, TU Bergakademie Freiberg, for enlightening discussions on the site geology, to Ms. Margot Schwab, Landeshauptstadt Dresden, for access to the "Hoher Stein" surface site, and to Ms. Susann Riechel, Staatsbetrieb Geobasisinformation und Vermessung Sachsen, for kindly providing the DGM1 geodata. -- This work was supported in part by the Helmholtz Association (NAVI VH-VI-417 and ERC-RA-0016), DFG (BE4100/4-1), and the COST Association (ChETEC \linebreak {CA16117}).


\end{document}